%% file: main.tex
\documentclass[runningheads]{llncs}
\usepackage{graphicx}
\usepackage{times}
\usepackage{helvet}
\usepackage{courier}
\usepackage{subcaption}
\usepackage{graphicx}
\usepackage{tabularx}
\usepackage[noend]{algorithmic}
\usepackage{algorithm} 
\usepackage{color,soul}
\usepackage{url}
\usepackage{hyperref}
\usepackage{wrapfig}

\begin{document}

\title{On Measuring the Diversity of Organizational Networks}
\author{Zeinab S. Jalali\inst{1} \and
Krishnaram Kenthapadi\inst{2} \and
Sucheta Soundarajan\inst{1}}
\authorrunning{Z. Jalali et al.}
\institute{Syracuse University, Syracuse NY, USA
\email{\{zsaghati,susounda\}@syr.edu}\\
\and
Amazon AI, Sunnyvale, CA,USA
\email{kkenthapadi@gmail.com}}

\maketitle          
\input{Files/Abstract.tex}

\input{Files/Introduction}

\input{Files/RelatedWork}
\input{Files/Problem}

\input{Files/Method}

\input{Files/Experiments}

\input{Files/Results}

\input{Files/Limitation}

\bibliographystyle{plain}
\bibliography{bib}
\clearpage

\end{document}

%% file: Files/Abstract.tex
\vspace{-0.2cm}
\begin{abstract}
\vspace{-0.3cm}
The interaction patterns of employees in social and professional networks play an important role in the success of employees and organizations as a whole. However, in many fields there is a severe under-representation of minority groups; moreover, minority individuals may be segregated from the rest of the network or isolated from one another.
While the problem of increasing the representation of minority groups in various fields has been well-studied, diversification in terms of numbers alone may not be sufficient: social relationships should also be considered.  In this work, we consider the problem of assigning a set of employment candidates to positions in a social network so that diversity and overall fitness are maximized, and propose \texttt{Fair Employee Assignment (FairEA)}, a novel algorithm for finding such a matching.  The output from \texttt{FairEA} can be used as a benchmark by organizations wishing to evaluate their hiring and assignment practices.
On real and synthetic networks, we demonstrate that \texttt{FairEA} does well at finding high-fitness, high-diversity matchings.

\vspace{-0.3cm}
\end{abstract}

%% file: Files/Introduction.tex
\vspace{-0.4cm}
\section{Introduction}
\vspace{-0.2cm}
In order for commercial and non-profit organizations to succeed, it is important for those organizations to recruit a workforce that is not only skilled, but also diverse, as diversity has been positively associated with performance~\cite{mazzola2020recruiting}.  However, diversity cannot be measured only in terms of numbers: it is known that negative effects may happen when a network is structured in a way that resources are not accessible through the social capital accessible to members of a minority group \cite{fung2015network}. Social capital consists of bridging resources from outside of an individual's group (inter-group connections) and bonding resources from internal group connections (intra-group connections)~\cite{lin2008network}.  

The literature contains a number of metrics for measuring network diversity/seg- regation, the most prominent being assortativity~\cite{newman2003mixing}.  However, when dealing with dynamic networks where new nodes are being added, it is useful to know not only what the diversity of a specific network snapshot \textit{is} after those nodes are added, but how good it \textit{could have been}.  In other words, if new nodes join a network, what is the best assortativity that one could possibly achieve, given pre-existing structure of the network and restrictions on where the new nodes can join?  

Our work is motivated by the example of an organization that is evaluating their hiring and employee assignment practices with respect to the diversity (gender, race, 
etc.) of the organizational network.  When positions are open, some set of candidates apply for those positions.  Each candidate has some amount (possibly zero) of suitability for each of the open positions.  If one's goal is to minimize segregation in the network while ensuring that each position is filled by a candidate who is suitable for that position, which candidate should one hire for each position?  If there are significant gender disparities in applications across job categories (e.g., if software engineer candidates are disproportionately male), then it may not be possible to achieve perfect diversity in hiring and assignment; but nonetheless, it is useful to know how well one can do.  One can imagine similar examples for, say, new graduate students joining an existing scientific collaboration network.

There has been a great deal of recent interest in fairness of hiring/assignment procedures (e.g., the Rooney Rule used by the American National Football League~\cite{collins2007tackling}).  This is because one cannot simply eliminate an existing professional network and replace it with a diverse network; and moreover, at the hiring stage, the candidate pool may itself be non-diverse or exhibit correlations between protected attributes and skillsets.  

In this work, we present \texttt{FairEmployeeAssignment (FairEA)}, a novel algorithm for determining how to assign a set of attributed node \textit{candidates} to \textit{open positions} in a network such that both the \textit{fitness} of the match between nodes and open positions and the \textit{diversity} of the resulting network are maximized.  We experimentally demonstrate that \texttt{FairEA} outperforms baseline strategies at achieving these goals. 

We note that in the United States and other countries, it is illegal to make employment decisions based on protected attributes like race or religion.\footnote{\url{https://www.eeoc.gov/laws/practices/}}  For this reason, although the output of \texttt{FairEA} is a matching of candidates to positions, it is \textit{not} intended to be used directly to make assignments.  Rather, the matching should be used as a baseline, and compared with the organization's actual practices in employee hiring and assignment, to assess the quality of an organization's assignment practices with respect to the diversity of the organizational network.

%% file: Files/RelatedWork.tex
\vspace{-0.2cm}
\section{Related Work}
\vspace{-0.2cm}
The recent scientific literature contains many studies on modeling bias in human recruiting systems.
For example, \cite{sgoutas2016we} examines strategies for hiring diverse faculty in universities, \cite{bjerk2008glass} shows the different likelihoods of hiring and promotion for candidates from different groups with equal skills, and \cite{newman2009recruitment} addresses the tradeoff between performance goals and company diversity.
\cite{mazzola2020recruiting} shows a positive relationship
between board member racial and gender diversity to performance of nonprofits.
Unfortunately, automating recruiting systems will not necessarily solve problems of discrimination in hiring~\cite{dobbe2018broader}. One solution is to make sure that protected attributes do not influence algorithmic decisions, but \cite{de2019bias} shows that gender bias exists even after scrubbing gender indicators from a classifier.  While traditional approaches measure diversity of organizations in terms of numbers~\cite{o2006beyond}, organizations are social networks, and network factors influence entrepreneurial success, mobility through occupational ladders, and access to employments~\cite{portes1998social}.

\begin{figure*}[t]
\centering
\includegraphics[scale = 0.23]
{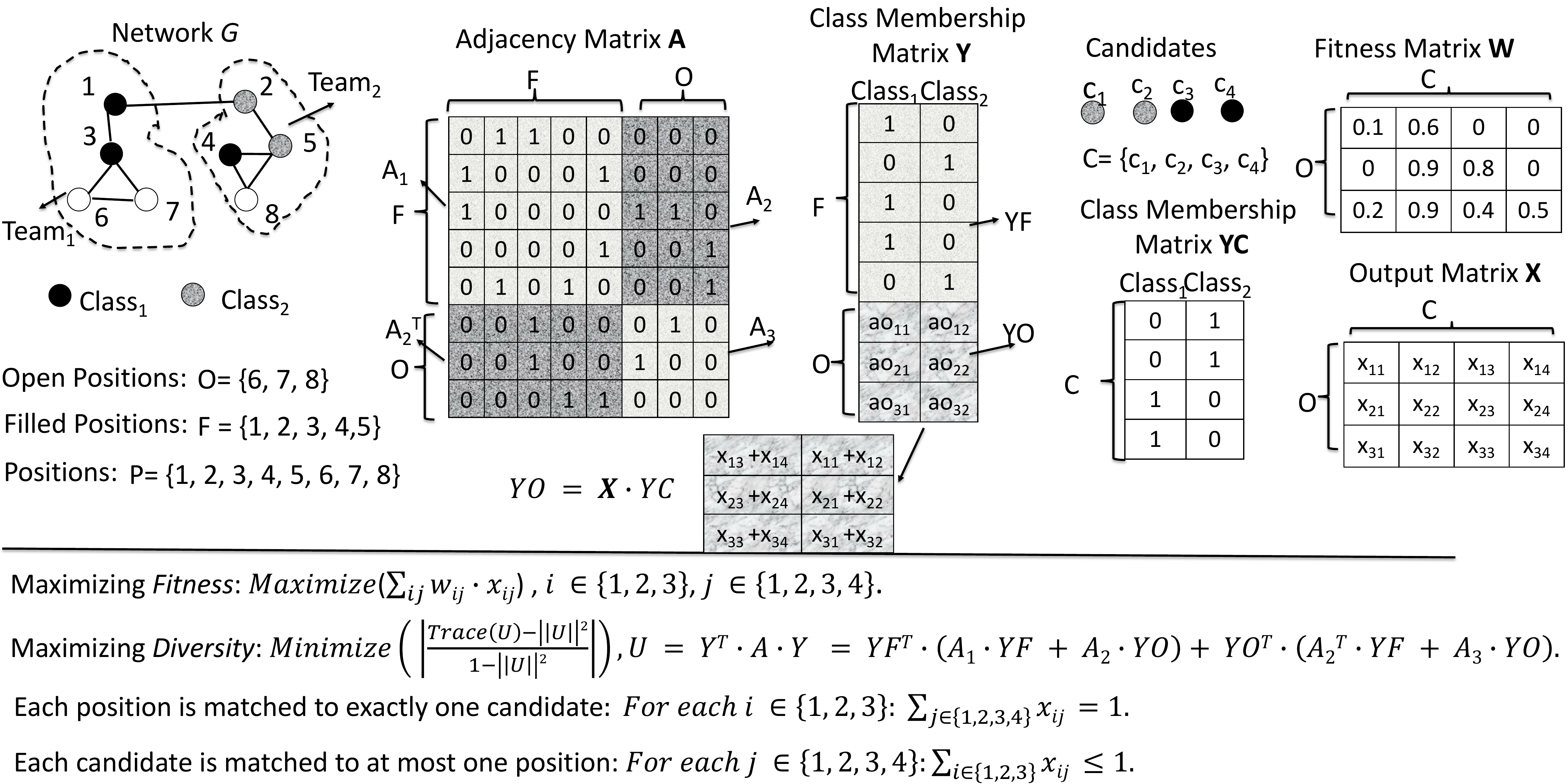}
\vspace{-0.1cm}
\caption{Overview of the problem.
}
\label{fig:ea}
\vspace{-0.5cm}
\end{figure*}

%% file: Files/Problem.tex
\vspace{-0.3cm}
\section{Problem Formulation}\label{sec:problem}
\vspace{-0.2cm}
We formulate this problem as a multi-objective problem in which the goal is to assign a set of newly-hired employees/employment candidates (without loss of generality, the `candidates') to open positions so as to maximize (1) the \textit{fitness} of employees to positions and (2) the \textit{diversity} of the organizational network, 
under the constraint that all open positions must be filled.  We compute diversity as assortativity, which measures the extent to which `like connect to like'. 
Figure~\ref{fig:ea} shows an overview of the problem.

The input for this problem consists of the following: 

(1) An undirected network $G = (P, E)$, representing the professional network of an organization.  Nodes represent positions ($s$ filled, $m$ open). Each edge $(p_i, p_j)$ represents either a real or expected professional interaction between the employees who currently fill or will fill positions $p_i$ and $p_j$ (i.e., those employees do interact, or are expected to interact once the positions are filled).  
 
(2) A set of $t$ candidates ($t \geq m$).  If $t = m$, this problem represents the case where new employees have already been hired and need to be assigned -- e.g., newly hired software engineers are being assigned to teams. If $t > m$, this problem can be viewed as a combination of the hiring and assignment problems.
 
(3) The fitness of each candidate $c_j$ for each position $o_i$ (how well-qualified $c_j$ is for $o_i$).  We assume that it is possible to match candidates to open positions such that each open position is filled subject to having at least one candidate with greater than zero fitness for each open position.
 
(4) An attribute of interest, such as gender, that divides employees/candidates into $k$ classes of attributes: $class_s,.., class_k$. We assume that this attribute is categorical and each node can be member of just one class (e.g., minority and majority).

The output is a matching of candidates to open positions. We refer to the input and output in the rest of the paper as described in Table~\ref{tab:notation}.

\begin{table}[t]
\footnotesize
\centering
\caption{Notation}
\label{tab:notation}
\begin{tabularx}{\textwidth}{l|l}
\bf {Symbol} & \bf {Definition} \\
\hline
\bf {$G(P,E)$} & {Unweighted, undirected attributed graph } \\
 \bf { $O = \{o_1,.., o_m\}$}& {Set of open/unfilled positions} \\
 \bf {$F = \{f_1,.., f_s\}$} & {Set of filled Positions}\\
 \bf { $P = \{p_1, .., p_n\}$} & {Set of positions (nodes of network $G$) },($F \cup O = P$)\\
 \bf {$Q = \{q_1,.., q_s\}$}& {Set of current employees($q_1,.., q_s$ fill
 $f_1,.., f_s$)} \\
 \bf {$ C = \{c_1,.., c_t\}$}& {Set of candidates to fill open positions}\\
 \bf{$\mathbf {W}_{m\times t}$}& {Fitness matrix,  $w_{ij}$:fitness of candidate $c_j$ for position $o_i$} \\
  \bf{$\mathbf{A}_{n \times n}$}& {{Adjacency matrix of network $G$}} \\
  \bf{$\mathbf{A_1}_{s \times s}, \mathbf{A_2}_{s \times m}, \mathbf{A_3}_{m \times m}$}& {Sub-matrices of $\mathbf{A}$, ($F$to $F$, $F$ to $O$, $O$ to $O$ edges)}\\
   \bf{$\mathbf {Y}_{n\times k}$}& { Membership matrix of current/future employees to $k$ classes} \\
\bf{$\mathbf {YF}_{s\times k}$,$\mathbf{YO}_{m\times k}$}& {Sub-matrices of $\mathbf{Y}$, Membership of current and future employees} \\
  \bf{ $\mathbf{YC}_{t\times k}$}& { Membership matrix of candidates to $k$ classes} \\
  \bf{$\mathbf X_{m\times t}$} & {binary output matrix, $x_{ij} =1$ if  $c_j$ is assigned to $o_i$}\\
   \bf{$\mathbf U_{k \times k}$}& { $u_{ij}:$ fraction of $class_i$ to $class_j$ edges} \\
\end{tabularx}
\vspace{-0.3cm}
\end{table}
\vspace{-0.3cm}
\subsection{Objectives}
\vspace{-0.1cm}
\texttt{FairEA} attempts to solve a multi-objective optimization problem with fitness and diversity-related objectives, described below.
\vspace{-0.5cm}

\subsubsection{Maximizing Fitness}\label{sec: fit}
The first goal is to maximize the \textit{fitness} of the assignment, subject to the constraint that all positions are filled, and each candidate fills at most one position.  This objective corresponds to the organization's primary goal of recruiting employees with the required skill sets~\cite{craig2015cost}.
 We compute the overall \textit{fitness} of a matching by summing the fitness of each pair of matched open positions and candidates. Higher values
indicate a matching of better qualified candidates for open positions.
This goal can be formulated as an optimization problem: 
$max$ $f_1 = \sum_{ij} w_{ij}\cdot x_{ij}$, such that for each  $1 \leq i \leq m$, $\sum_{1 \leq j \leq t} x_{ij} =1$ and for each $1 \leq j \leq t$, $\sum_{1 \leq i \leq m} x_{ij} \leq 1$.

\vspace{-0.3cm}
\subsubsection{Maximizing Diversity}
We compute \textit{diversity} as \textit{assortativity}, which measures the extent to which a minority group is integrated into the larger network based on the number of inter- vs intra-group connections.  If we consider matrix $\mathbf{U}$ with $u_{ij}$ to be the fraction of edges in the network that connect a node from class $i$ to class $j$, the assortativity coefficient is equal to $\frac {Trace(\mathbf{U}) - ||\mathbf{U}^2||}{ 1 - ||\mathbf{U}^2||}$ where $||\mathbf{U}^2||$ means sum over all elements in $\mathbf{U}^2$~\cite{newman2003mixing}. Positive values show more intra-group connections and negative values show more inter-group connections in network.  Our goal is to minimize the absolute value of assortativity, so that groups are neither segregated nor isolated.
Let $\mathbf{A}_1$, $\mathbf{A}_2$, and $\mathbf{A}_3$ be sub-matrices of adjacency matrix $\mathbf{A}$, where $\mathbf{A}_{1(s\times s)}$ shows the existence of the edges between nodes in $F$,
$\mathbf{A}_{2(s\times m)}$ shows the existence of the edges from nodes in $F$ to nodes in $O$,
and $\mathbf{A}_{3(m\times m)}$ shows the existence of the edges between nodes in $O$. Let $\mathbf{YO}_{m\times k}=\mathbf{X}\cdot \mathbf{YC}$ be the binary matrix of the membership of candidates that will fill the open position to each attribute class. Then we have:    $\mathbf{U} =\mathbf{Y}^T \cdot \mathbf{A} \cdot \mathbf{Y}$
$= \mathbf{YF}^T \cdot (\mathbf{A}_1 \cdot \mathbf{YF} + \mathbf{A}_2 \cdot \mathbf{YO}) + 
 \mathbf{YO}^T \cdot( \mathbf{A}_2^T \cdot \mathbf{YF} +\mathbf{A}_3 \cdot \mathbf{YO})$ and {this goal can be formulated as an optimization problem as:} $min$ ~~ $f_2 =| \frac {Trace(\mathbf{U}) - ||\mathbf{U}^2||}{ 1 - ||\mathbf{U}^2||}|$.

\vspace{-0.1cm}
\vspace{-0.3cm}
\subsubsection{Other Constraints:}  
In certain cases, there may be other constraints that one wishes to consider.  For example, if a minority group is very small, a low-homophily network would indicate that members of that group are isolated.  Organizations may want to avoid such an outcome, because such isolated individuals may be unable to find other members of a group and form a \textit{community of support}, where peers share their experiences and discuss their problems and provide social support for each other~\cite{williams2017providing}.  In such cases, appropriate constraints (e.g., the minimum number of members from each protected group per team).  We give an example of this in the next section.

\vspace{-0.1cm}
\subsection{Challenges}\label{sec:challenges}
\vspace{-0.1cm}
First, note that the problem considered here is NP-hard via a reduction from \texttt{Unweighted Max Cut}~\cite{grotschel1981weakly}.  Second, the problem of minimizing \textit{diversity} is not convex, and is neither sub-modular nor super-modular. Third, although the problem can be formulated as an integer program, this process is computationally slow. 
These challenges suggest that even a fast approximation algorithm may not exist.  As such, we present a heuristic and demonstrate its strong performance experimentally.

%% file: Files/Method.tex
\vspace{-0.2cm}
\section{Method} \label{sec:method}
\vspace{-0.2cm}
We propose \texttt{FairEmployeeAssignment (FairEA)}, a method for assigning candidates to positions with the goals of maximizing \textit{fitness} and \textit{diversity}.
Assume that we are given input and desire output as described in Section~\ref{sec:problem}. In our initial discussions, we assume that candidates are divided into $k=2$ classes, and in Section~\ref{sec:mod}, we explain how \texttt{FairEA} can be generalized for $k>2$.

\texttt{FairEA} consists of a sequence of iterations, where each iteration $i$ consists of the following three steps:
\begin{enumerate}
 \item Select two subsets $O_i$ from $O$ and $C_i$ from $C$ using the \textit{FairEA selection process} described later in this section.
 \item Assign $C_i$ to $O_i$ using the \textit{FairEA Matching Process}, as described later in this section.
 \item If $|O_i| < |O|$ then increment $i$ and return to (1). Otherwise, terminate.
\end{enumerate}

\vspace{-0.6cm}
\subsubsection{\texttt{FairEA} Selection Process} \label{sec: fsp} 
For each pair of open positions $o_a \in O$ and candidates $c_b\in C$, where $w_{ab}>0$, consider two scores: the \textit{fitness score}, given by $w_{ab}$, and the \textit{diversity score}. The diversity score is defined as 1 if $c_b \in class_j$ ($j \in \{1,2\}$), and the number of positions adjacent to $o_a$ filled by an employee from $class_j$ is less than the number of positions adjacent to $o_a$ filled by an employee from the other class. 

Using these two scores, we use the Pareto Optimality technique described in~\cite{censor1977pareto} to select subsets from  $O$ and $C$.  At each iteration $i$ that the \textit{selection process} is called, the output contains all the open positions and candidates that are present in the pairs appearing in the top $i$ Pareto front sets. 

A \textit{Pareto front set} consists of all points that are not dominated by any other point (a point $(x_1, y_1)$ is dominated by $(x_2, y_2)$ if $x_2 > x_1$, $y_2 > y_1$, or $x_2 \geq x_1$, $y_2 > y_1$, or $x_2 > x_1$, $y_2 \geq y_1$).  To find the top $i$ Pareto front sets, one finds the first Pareto front set as just described, removes all selected points, and repeats for $i$ iterations.  

\vspace{-0.5cm}
\subsubsection{\texttt{FairEA} Matching Process} \label{sec: fmp}
The \texttt{FairEA} matching process is based on the augmenting path approach from the Hungarian algorithm for weighted bipartite matching~\cite{grinman2015hungarian}:
\begin{enumerate}
 \item Generate a bipartite graph $B$ from $C_i$ to $O_i$, where edges represent each pair of qualified candidates $o_a \in O$ and open positions $c_b\in C$ $\&$ $w_{ab}>0$. To set weights:
 \begin{itemize}
\item Sort the fitness score and diversity score computed, as described earlier.
\item Set edge weights based on their position in the Pareto front sets levels.
If $(o_a, c_b)$ is present in level $l$, its score is equal to $\frac{1}{l}$.
 \end{itemize}
 \item Create a labeling $l$ ($l[j]=0$ for each node $v_j$ in $B$), an empty matching $M$, and an empty bipartite graph $B_l$.
 \item If all elements in $O_i$ are matched and present in $M$, update matrix $\mathbf X$ based on matching $M$ and stop. Otherwise, update labeling $l$ and bipartite graph $B_l$.
 \begin{itemize}
\item For each unmatched open position $o$ in graph $B$, set $l(o)$ to the maximum weight of edges connected to node $o$ in $B$.
\item For each matched open positions $o$ in $M$ (if any), if $o$ is matched to $c$, $l(o) = weight(c, o)$).
\item For each candidate $c$ in graph $B$, set $l(c)$=0.
\item Graph $B_l$ contains edges $(o,c)$ where $o_a \in O_i$, $c \in C_i$, $l[o] + l[c] \leq weight(o,c)$.
 \end{itemize}
\item Pick an unmatched open position $o_a \in O_i$. Let $S=\{o_a\}$ and $T=\{\}$ and
 \item Let $N(S)$ be the set of neighbors of nodes
 from $S$ in $B_l$. 
 \item If $N(S) = T$, update labels:
 \begin{itemize}
\item $\alpha = min_{o \in S, c \notin T} {l(o)+l(c)-w(o,c)} $,

$l(o) = l(o)- \alpha$, if $o \in S$,  $l(c) = l(c) + \alpha$, if $c\in T$.
 \end{itemize}
 \item If $N(S) \neq T$, pick $c \in N(S) - T$.
 \begin{itemize}
\item If $c$ is not matched find an augmenting path from $o_a$ to $c$. Augment $M$ and update $G$ based on the new assignments. Update weights of edges in $B$ using the same approach described in (1) and return to (2).
\item If $c$ is matched to $o_b$, extend the alternating tree. Add $o_b$ to $S$ and add $c$ to $T$ and return to (4).
 \end{itemize}

\end{enumerate}

In each iteration, the matching is improved and $G$ is updated. For newly assigned positions, the attribute will be the attribute of the matched candidate.
\vspace{-0.2cm}
\subsection{Handling Constraints}\label{sec: constraint}
\texttt{FairEA} can handle other diversity-related goals in the form of constraints. As an example, we consider the constraint that no minority individual should be isolated from other minority individuals. To address this (or any) constraint, add a step to \texttt{FairEA}.

Suppose a company has $k'$ disjoint teams and wants to ensure that minorities from each team $i$ are grouped with at least $t_i$ other members of that minority group.
In this step, \texttt{FairEA} assigns a set of best qualified candidates from each class $j$ to open positions in the team $i$ that has fewer than the threshold $t_i$ 
employees from class $i$. (In practice, this can be accomplished via cluster hiring). A threshold of 0 indicates that avoiding isolation is not necessary. In this step, after each matching, \texttt{FairEA} ensures that all the remaining open positions can be filled (i.e., there is at least one distinct candidate with fitness function greater than zero for each remaining open position).  It may sometimes not be possible to reach the threshold for a specific team (e.g., there are not enough open positions in the team). In such cases, so as to not \textit{prevent} assignment of a qualified minority to such a team, the algorithm can perform the assignment, but notify the organization. This information can then be used by the organization: e.g., individuals who lack access to other minorities can be enrolled in a mentor/mentee program.

Specifically, \texttt{FairEA} performs the following: For each team $i$, denote the number of individuals from $class_j$, $j \in \{1,2\}$ as $|c_{ji}|$, where $|c_{ji}|< t_i$. Sort all pairs of $\{(o_a,c_b), o_a \in O$ and $o_a \in class_j$ and $c_b\in C \}$ based on $w_{ab}$ (fitness of $c$ for $o$) in descending order. Next, iterate over all the elements $(o_a,c_b)$ in the sorted set. If both $o_a$ and $c_b$ are not already matched (i.e., all elements of row $a$ and column $b$ in matrix $\mathbf{X}$ are zero) and there is at least one possible complete matching from remaining candidates to remaining open positions, set $x_{ab}=1$ and remove matched elements from $O$ and $C$. Continue the iteration until $t_i - |c_{ji}|=0$ (sufficient new matchings are established) or there are no elements in the set. In the end, if $|c_{ji}|$ is still less than $t_i$, return the team $t_i$ for notifying the organization.

\vspace{-0.2cm}
\subsection{Variations on FairEA}\label{sec:mod}
\vspace{-0.1cm}
It is easy to modify \texttt{FairEA} for other settings:

\textit{Non-binary attributes:} If the protected attribute has more than two classes, the diversity score calculation has to be changed.  For each pair of qualified candidates $o_a \in O$ ($o_b \in class_j$) and open positions $c_b\in C$, where $w_{ab}>0$, the diversity score is defined as  $\frac{ct- ca}{ct} $, where $ct$ is the total number of filled positions adjacent to $c_b$ and $ca$ is the number of filled positions adjacent to $c_b$ that filled by an employee from $class_j$.

\textit{Multiple attributes of interest:} 
If there are multiple attributes of interest (e.g., race and gender), by combining them into one new attribute we can address the problem using \texttt{FairEA} for $k$ classes. For instance, suppose we have $k_1$ classes for gender and $k_2$ classes for race, we can generate a new attribute called identity with at most $k_1*k_2$ classes. Because this may be a large number of combinations, these intersectional classes may be merged as appropriate.

%% file: Files/Experiments.tex
\vspace{-0.2cm}
\section{Experimental Setup}\label{sec:experiments}
\begin{wraptable}{r}{5cm}
\scriptsize
\centering
\vspace{-0.8 cm}
\caption{Dataset statistics. }
\vspace{-0.2 cm}
\label{table:dataset}
\begin{tabular}{l|l|l| l| l} 
\bf{Name} & \bf{\#nodes} & \bf{\#edges} & \bf{ Assort.}
& \bf{  Attributes} \\
\bf{} & \bf{} & \bf{} & \bf{~~~Coeff.} &\bf{ (Maj, Min)} \\ \hline
\bf{CC(M)} & 46& 552 &  -0.02
& (77\%, 23\%) \\ 
\bf{CC(H)} & 46& 552 &  0.37
& (57\%, 43\%) \\ 
\bf{RT(M)} & 77& 1341 & 0.02
&(88\%, 12\%)\\
\bf{RT(H)} & 77& 1341 & 0.43 
&(65\%, 35\%)
\\ \hline

\bf{Nor(L)} & 1522& 4143 & -0.19 
&(61\%, 39\%) \\ 
\bf{Nor(M)} & 1091& 3418 & .08 
&(90\%, 10\%) \\ 
\bf{Nor(H)} & 1421& 3855 & .29 
& (64\%, 36\%) \\ 
\hline 
\bf{FO(H)} & 288 & 2602 &.86
&(70\%, 30\%) \\ 
\bf{DO(H)} &265 &921 &0.92
& (70\%, 30\%) \\ 
\hline 
\bf{SF(L)} & 1000&4000& -0.30 
&(69\%, 31\%) \\ 
\bf{SF(M)} & 1000&4000& 0.07 
&(69\%, 31\%) \\ 
\bf{SF(H)} & 1000&4000& 0.39 
&(69\%, 31\%) 
\end{tabular}
\vspace{-0.8cm}
\end{wraptable}
In our experiments, we first demonstrate that \texttt{FairEA} does well at matching employees to positions (with respect to fitness and diversity).  Second, we provide an example of \texttt{FairEA} on a real-world organizational network.\footnote{For replication, we have posted our code and data at \url{https://github.com/SaraJalali/FairEmployeeAssignment}.}

\vspace{-0.2cm}
\subsection{{Datasets}}
\vspace{-0.1cm}

It is difficult to get real data for this problem- in particular, on candidate pools and demographics.  Thus, to test \texttt{FairEA} under a wide array of conditions, we use both real and synthetic network topologies and attributes and simulate candidate pools.  Statistics of datasets are shown in Table~\ref{table:dataset}.  (L, M, H) notations  indicate low, medium, and high values of assortativity (segregation) in the network.

\subsubsection{Real Network Topologies}
First, we use the Norwegian Interlocking Directorate network (Nor), which describes connections among directors of public companies in Norway{\footnote{\url{http://www.boardsandgender.com/data.php}}}. This dataset includes the `gender' attribute.  We selected two snapshots of this network, the first from February 2003 (\textbf{Nor(M))} and the second from October 2009 (\textbf{Nor(L)}).  These networks have, respectively, the highest and lowest levels of gender assortativity among all snapshots.  We also add synthetic attributes to the August 2011 to obtain a with high assortativity, denoted as \textbf{Nor(H)}. 

Next, we consider a set of intra-organizational networks showing interactions in a Consulting Company (CC) and a Research Team (RT),\footnote{\url{https://toreopsahl.com/datasets/}} and consider different attributes to generate high and medium levels of assortativity. \textbf{CC(M)} is obtained from the `gender' attribute, and \textbf{CC(H)} from the `location' attribute (`Europe' vs. `USA'). \textbf{RT(L)} uses the `tenure' attribute (less than one year vs. more than one year), and \textbf{RT(H)} uses the `location' attribute (`London' vs. the rest of Europe). (In practice, not all of these attributes are things we care about in the context of network diversity; they were chosen for purposes of demonstrating the algorithm.)  Additionally, these two sets of networks contain information about the the organisational level of employees, which we use in our second experiment.
\vspace{-0.3cm}
\subsubsection{Synthetic Networks}
To evaluate \texttt{FairEA} on additional network topologies, we construct synthetic datasets.  First, \textbf{FO} (functional organization) has 6 teams and 12 sub-teams with equal number of nodes in each team, and follows the FedEx organizational chart pattern~\cite{lumen}.
\textbf{DO} (divisional organization) has 3 divisions and 40 teams with an equal number of nodes in each team, and follows the Department of Energy organizational chart pattern~\cite{lumen}.
We add a synthetic binary attribute so that members of each team are from one class.  We next generated a set of power-law (SF) networks~\cite{holme2002growing}. We generate three networks with the same topology, and assign attributes to have low \textbf{SF(L)}, medium \textbf{SF(M)}, and high \textbf{SF(H)} assortativity levels.

\vspace{-0.3cm}
\subsection{Open Positions, Teams, and Candidate Pool}\label{sec:OTC}
\vspace{-0.1cm}
For each network, we run 100 trials. In each trial, we sample $10\%$, $20\%$, and $30\%$ of nodes randomly as open positions.
To simulate the pool of candidates we consider two cases: (1) the candidate pool consists of the nodes set to open (with the same attributes), 
and (2) the candidate pool consists of two copies of each node set to open. 
The first setting corresponds to the case where a `batch' of new employees has been hired, and now the employees need to be assigned to teams without considering the hiring process.
The second setting corresponds to the case where we consider both hiring and assignment procedures. Moreover, the way we assign attributes to nodes ensures that changes in homophily are actually due to employee \textit{assignment}, rather than changes in attributes.

\vspace{-0.2cm}
\subsection{Fitness Functions}\label{sec:ff}
\vspace{-0.1cm}

The fitness function governs which candidates are suitable for which positions. 
For the first sets of experiments- the evaluation of \texttt{FairEA}-
we consider two fitness functions.  In $F_1$, candidates are qualified for four randomly selected positions with fitness equal to a random number in $(0, 1)$.  In $F_2$, candidates are fit for the four open positions closest to the position that the candidate had previously filled 
with fitness equal to a random number in $(0, 1)$.

\vspace{-0.3cm}
\subsection{Baseline Methods}
\vspace{-0.1cm}
We use three baseline methods: (1) \texttt{Random}, which randomly assigns qualified candidates to each open position; (2) The weighted \texttt{Hungarian} algorithm, where the input is a bipartite graph whose two sides correspond to open positions and candidates. An edge $(o_a,c_b)$ exists if $w_{ab}>0$, and the weight of this edge is the sum of $w_{ab}$ and the diversity score as described in section~\ref{sec: fsp}; and
(3) \texttt{Optimization}, which uses the \texttt{IPOPT} solver in the GEKKO optimization suite~\cite{beal2018gekko} for solving the optimization problem with the two goals of maximizing \textit{fitness} and \textit{diversity}. This is the simplified version of the problem where \textit{fitness} is maximized, as described in section~\ref{sec: fit} and \textit{diversity} is optimized by decreasing the gap between number of neighbors from $class_i$ to number of neighbors from $class_j$ for each newly assigned position.

%% file: Files/Results.tex
\vspace{-0.2cm}
\subsection{Metrics}
\vspace{-0.1cm}
 We report results using the following metrics:
 \vspace{-0.1cm}
\begin{itemize}
 \item The \textit{overall fit score} is the sum of the fitness scores for each matching. Let $FS_h$ and $FS_l$ be \textit{overall fit score} of the best and worst possible matching in terms of fitness of employees for the open positions respectively and $FS_a$ be the \textit{overall fit score} of the network $G$ after assignment using desired method.  Then we define \textit{Percentage Improvement in Fitness}$= \frac{FS - FS_l}{FS_h- FS_l}\cdot 100$.
 
\item The \textit{diversity} of the network is measured by the \textit{assortativity coefficient}~\cite{newman2003mixing}. Let $AC_b$ be the \textit{assortativity coefficient} of $G'$, the subgraph of initial network $G$ consisting only of filled positions, and $AC_b$ be the \textit{(assortativity coefficient} of the network $G$ after assignments are made. Then the \textit{Percentage Improvement in Assortativity} $=\frac{|AC_b| - |AC_a|}{|AC_b|}\cdot 100 $.
 \item The \textit{fraction of minorities} in team $i$ is  $FM_{i}=\frac{min(|c1_{i}|,.., |ck_{i}|)}{|c1_i|+... |ck_i|}$ where $|cj_i|$ is the number of individuals from $class_j$ in team $i$. \textit{Isolation Score} is the average fraction of minorities.
 \textit{Isolation Score}$=\frac{1}{k} \cdot \sum_{1 \leq i \leq k}{FM_{i}}$.

\end{itemize}
\vspace{-0.1cm}
\section{Results and Analysis}\label{sec:res}
\vspace{-0.2cm}
We first compare \texttt{FairEA} to the baseline algorithms in order to evaluate its performance algorithmically.  Next, we use real intra-organizational networks to demonstrate how \texttt{FairEA} might be used in practice.

 \begin{figure*}[t]
\centering
\includegraphics[scale = 0.33]
{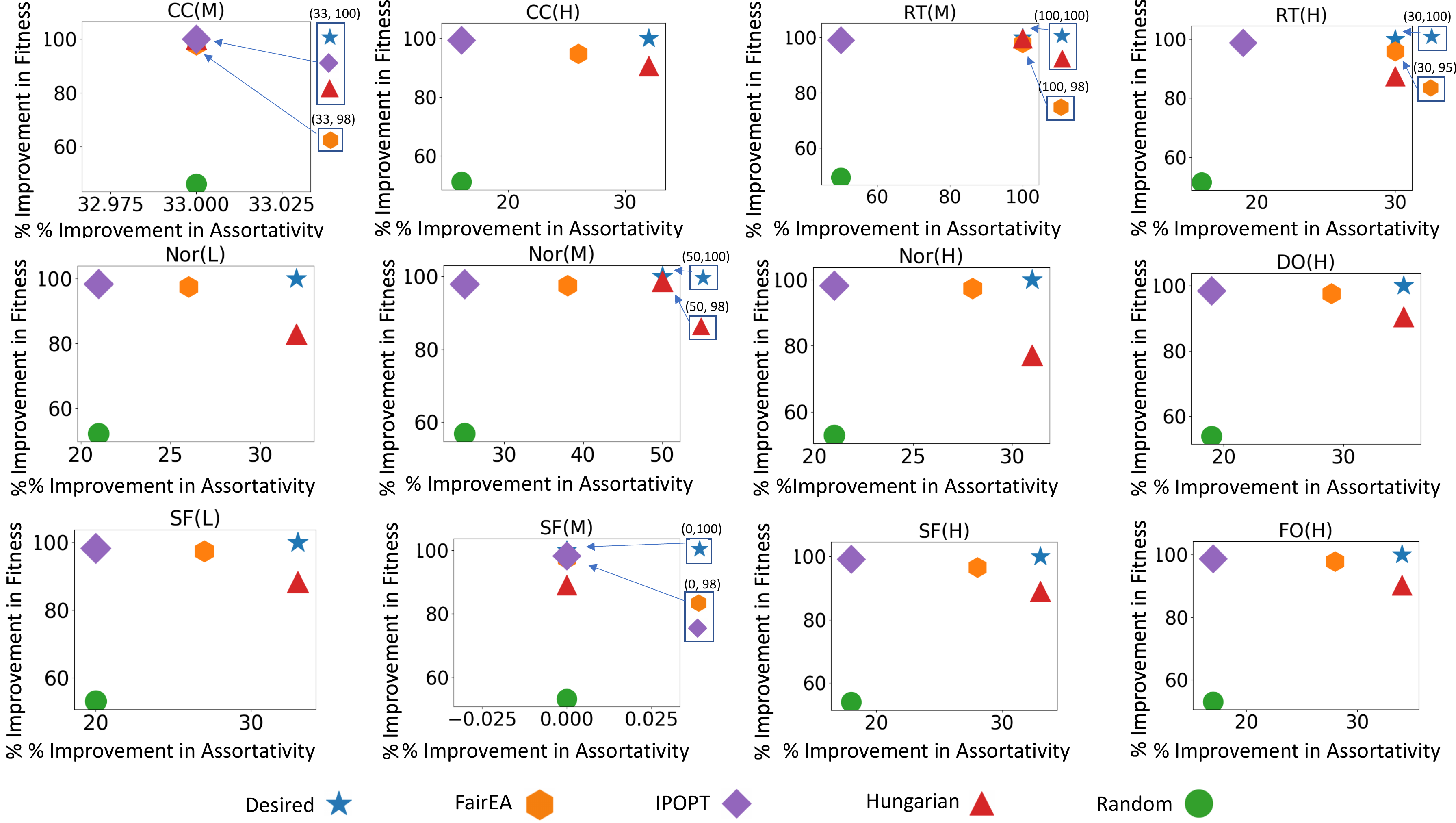}
\vspace{-0.3cm}
\caption{Comparison results of percentage improvement in fitness and assortativity for FairEA and baseline methods.  The ideal solution lies in the top right corner. \texttt{Hungarian} is good at increasing \textit{diversity} and \texttt{IPOPT} is good at increasing \textit{fitness}, but \texttt{FairEA} is good at increasing both. \texttt{Hungarian} performs well when assortativity is low (network is diverse).
}
\label{fig:exp1_f1}
\vspace{-0.5cm}
\end{figure*}

\vspace{-0.2cm}
\subsection{\texttt{FairEA} Evaluation}
\vspace{-0.1cm}
Here, we compare \texttt{FairEA} and baseline algorithms with respect to \textit{diversity} and \textit{fitness}.  Figure~\ref{fig:exp1_f1} shows results for \textit{percentage improvement in fitness} and \textit{percentage improvement in assortativity}, where the number of candidates is equal to the number of open positions, with fitness function $F_1$ (candidates are qualified for positions across the network) and $10\%$ open positions. The ideal solution (high fitness and diversity) is in the top right depicted as a star.  Results for fitness functions $F_2$ (candidates are qualified for positions in a specific area) are similar.

In most cases, results for \texttt{FairEA}, \texttt{IPOPT} and \texttt{Hungarian} are in a non-dominated set, with results of \texttt{FairEA} having the lowest crowding distance. More simply, we see that \texttt{Hungarian} does very well with respect to \textit{diversity} (especially when the network was already diverse), \texttt{IPOPT} does very well with respect to increasing \textit{fitness}, and \texttt{FairEA} does well at increasing both.  

To summarize results, we compute the \textit{average percentage improvement in fitness} and \textit{average percentage improvement in assortativity} over all datasets with high, medium and low levels of assortativity for each method.  When the size of the candidate pool is equal to the number of open positions, \texttt{FairEA} achieves at least $97 \%$ of the maximum fitness score while improving the assortativity coefficient value by $39\%$, $56\%$ and $67\%$ for $10\%$, $20\%$ and $30\%$ of open positions respectively. (Results were similar for other experimental settings.)
Overall, while \texttt{IPOPT} increases \textit{fitness}, it performs poorly on diversity.  This demonstrates  that simply considering the number of neighbors of a node from each class for newly assigned candidates is not sufficient.
\texttt{Hungarian} performs well when the number of open positions is small, but performance decreases as the number of open positions increases. In contrast, \texttt{FairEA} consistently does well.

\begin{figure*}[t]
\centering
\includegraphics[scale = 0.18 ]
{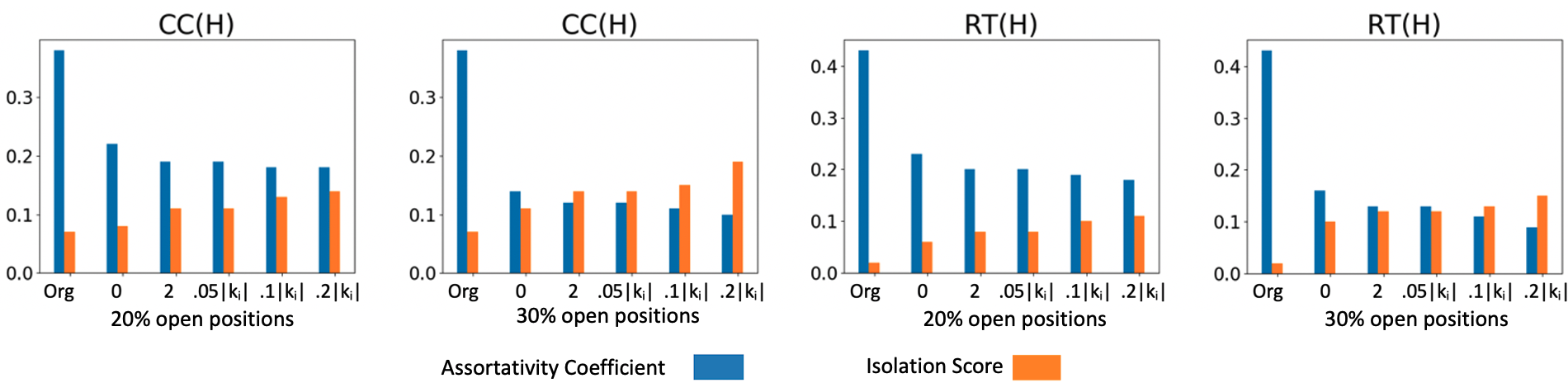}
\vspace{-0.2cm}
\caption{Results of assortativity coefficient and isolation score over original network (Org) and networks after assignment with different \textit{isolation} thresholds $t_i \in \{0, 2, 0.05\cdot|k_i| , 0.1\cdot|k_i|,0.2\cdot|k_i| \}$ where $|k_i|$ is size of team $team_i$. Both networks have the potential to become fair by at least $50 \%$. 
}
\vspace{-0.2cm}
\label{fig:exp3}
\vspace{-0.2cm}
\end{figure*}

\vspace{-0.3cm}
\subsection{Example Usage of \texttt{FairEA}}
\vspace{-0.1cm}

We next illustrate how \texttt{FairEA} can be used in practice- i.e., to evaluate an organization's hiring/assignment practices- on the intra-organizational networks CC and RT, which contain position level-related annotations.  In such a setting, the organization would identify the set of all positions that have been open in the recent past (whatever timespan is desired), and would use the actual applicants to those positions to form the candidate pool.  Because we do not have access to this data, we mark a random $p\%$ of the positions as open and consider the employees that fill those position as candidates.  We say that individuals are fit for positions at their level. 

Recall that in addition to optimizing for diversity and fitness, \texttt{FairEA} can accommodate constraints related to isolation (ensuring that minority individuals are not too far away from other minority individuals, which can have a negative effect on effectiveness~\cite{cohen1997makes}), and such constraints may affect performance with respect to fitness and diversity.  Here, in addition to evaluating fitness and diversity, we also evaluate the effect of such a constraint. 
We compute the \textit{Percentage Improvement in Fitness}, \textit{Percentage Improvement in Assortativity} and \textit{Percentage Isolation Score} of \texttt{FairEA}'s results when requiring that the number of minority group individuals in each group is at least $\{0, 2, 0.05\cdot|k_i| , 0.1\cdot|k_i|,0.2\cdot|k_i| \}$ where $|k_i|$ is size of team $team_i$.

We consider networks \textbf{CC(H)} and \textbf{RT(H)}, both of which are extremely segregated: the original networks have 
(Assortativity Coefficient \&  \textit{Isolation Score}) $(0.38 \& 0.07)$ and $(0.43 \& 0.02)$ respectively. Both of these networks are extremely segregated and have fewer than $10\%$ minorities in each team.

When applying \texttt{FairEA}, $20\%$ and $30\%$ of the positions are open, with different threshold levels for isolation, we see huge improvements in segregation. Figure~\ref{fig:exp3} shows the results of \textit{Assortativity Coefficient} 
and \textit{Isolation Score}: these large improvements in both assortativity and isolation indicate that both networks have great potential to become more fair.

%% file: Files/Limitation.tex
\vspace{-0.3 cm}
\section{Discussion, Limitations, and Conclusion}
\vspace{-0.1cm}
In this work, we proposed \texttt{FairEA}, a novel algorithm that can be used to gauge discrimination with respect to a protected attribute.  
Compared to baselines, \texttt{FairEA} does well at finding high-diversity, high-fitness matchings. While \texttt{FairEA} addresses an abstracted problem, it is a step towards a computational approach to create a diverse workplace in terms of social connections. 

This work is intended as a step towards remedying segregation and isolation in organizational networks by providing a simple approach to assessing the quality of hiring/assignment practices.  We acknowledge that there are substantially more considerations - such as the process used to generate the candidate pool- that go into evaluating hiring/assignment procedures than those described here, and hope that future work will build on what we have presented.